\documentclass[aps,prb,twocolumn,showpacs,superscriptaddress]{revtex4}

\usepackage{amsmath}
\usepackage{bm}
\usepackage{graphicx}

\begin{document}

\title{Heat transport and thermal rectification in quasi-one-dimensional systems.}

\author{E.\ D\'{\i}az}
\affiliation{Institute for Materials Science, Technische Universit\"at Dresden, 01062 Dresden, Germany} 
\affiliation{GISC, Departamento de F\'{\i}sica de Materiales, Universidad Complutense, E-28040 Madrid, Spain}
\author{R.\ Gutierrez}
\affiliation{Institute for Materials Science, Technische Universit\"at Dresden, 01062 Dresden, Germany} 
\author{G.\ Cuniberti}
\affiliation{Institute for Materials Science, Technische Universit\"at Dresden, 01062 Dresden, Germany} 
\affiliation{National Center for Nanomaterials Technology, POSTECH, Pohang 790-784, Republic of Korea}

\begin{abstract}
In this work we investigate heat conduction along a ladder-model conformed by 
two coupled one dimensional lattices with different anharmonicity. We study how
the interchain coupling modifies the thermal properties of the isolated systems. 
For a large enough coupling strength, we demonstrate that a harmonic lattice interacting with
an anharmonic one is able to support a linear thermal gradient when it is connected
to two heat reservoirs at different temperatures. 
We estimate this critical coupling by applying the self-consistent phonon theory (SCPT) to the anharmonic 
counterpart. 
By exchanging the heat baths connections between the harmonic and 
the anharmonic chains, our results show that the coupled system reveals as a thermal rectifier.
\end{abstract}

\pacs{
44.10.+i   
05.60.-k   
66.70.-f   
}
\maketitle
\section{Introduction}
\label{intro}
The current development of nanomaterials for molecular electronics 
has triggerd the study of heat conduction in low-dimensional systems from new perspectives.
It is well known that the derivation of the thermal properties in one and two dimensional systems 
is a controversial topic which has been deeply studied in the last two decades.~\cite{Lepri03,Dhar08}
In the case of perfect harmonic lattices, the lack of phonon-phonon interactions 
leads to no thermal resistivity, which gives rise to an infinite thermal conductivity in the thermodynamic limit.
This means, Fourier's law is not valid in these systems and heat conduction exhibits anomalous behavior.
Nonintegrability and an external substrate potential constitute the sufficient conditions to completely change
this scenario.~\cite{Hu98} 
In such a case, the conservation law of total momentum is violated and 
Fourier's law reveals as valid. Recently, the most studied proposals are 
the Frenkel-Kontorova (FK),~\cite{Gillan85,Tsironis99,Li06_fi4,Zhong09} and the $\phi^4$ potentials.~\cite{Aoki00,Hu99,Li07,Piazza09}
Several authors have 
revisited the unusual thermal properties of one dimensional systems
in order to propose different applications such as thermal rectifiers or thermal transistors.~\cite{Terraneo02,Hu06,Li06,Zhang10}
Some of these studies have been extended to two and three dimensional structures~\cite{Lee05, Saito10} and complex networks~\cite{Liu07,Liu10}, where it has been 
addressed the relevant influence of the coupling among several one dimensional lattices.

More interestingly, the analysis of the mechanisms mediating energy flow in low dimensional biomolecules
is a fundamental issue for the understanding of many biologically relevant
functions. The electronic and vibrational degrees of freedom of biomolecular systems, specially those containing helix structures
, i.e. double-helix DNA or $\alpha$-helices in proteins, can be modeled as ladder structures 
of coupled one-dimensional lattices.~\cite{Iguchi97,Diaz07,Scott92, Henning02} 
Recently, the thermal conductivity of double-stranded molecules has been studied in Ref.~\onlinecite{Zhong10} where
the interchain interaction between two identical chains has revealed as a positive or negative effect on heat conduction,
depending on the strength of the nonlinearity present in the system.

In this paper we address some interesting issues about heat transport along double-stranded molecular systems
which remain open. In particular, we consider a ladder-model where two different lattices, a harmonic and an anharmonic one, are 
coupled by harmonic forces. In view of the clearly different thermal properties of these two subsystems
when they are isolated, our work analyzes how the coupled system behavior is affected depending on the coupling strength.
In addition, due to the asymmetry present in the system, we address the possibility of heat rectification.  

\section{Model}
\label{model}
In this section we will present the theoretical formalism we will be dealing with 
in our study on heat transport along double-stranded molecules. 
In particular, we consider a system conformed by a harmonic one dimensional lattice 
which is coupled by harmonic forces to an anharmonic one. 
The dimensionless Hamiltonian of such a system is written as follows:
\begin{eqnarray}
{\cal H}&=& {\cal H}_{H}+{\cal H}_{A}+{\cal H}_{int} \ ,\nonumber \\
{\cal H}_{A}&=& \sum_{n=1}^N \frac{1}{2} \dot{x}^2_n + W(x_{n},x_{n-1})+V(x_n)\ ,\nonumber \\
{\cal H}_{H}&=& \sum_{n=1}^N \frac{1}{2} \dot{y}^2_n + W(y_{n},y_{n-1})\ ,\nonumber \\
{\cal H}_{int}&=& \sum_{n=1}^N k_{int} W(x_{n},y_{n})\ .
\label{Hamiltonian}
\end{eqnarray}
 
${\cal H}_{A}$ and ${\cal H}_{H}$ describe the dynamics of the subchains, the harmonic (H) and the anharmonic (A) one respectively, 
each one consisting of $n=1\ldots N$ sites. 
Both are affected by a harmonic potential $W(x_n,x_{n-1})=\frac{1}{2}(x_n-x_{n-1})^2$ and chain (A) is considered within the Frenkel-Kontorova model.
Thus, chain (A) is affected by an on-site potential $V(x_n)=\frac{-V_0}{4\pi^2}\cos 2\pi x_n$, whose strength is defined by $V_0$.
For the sake of simplicity, we will consider the same elastic constant for both chains, $k$, in terms of which the harmonic 
interchain coupling $k_{int}$ is expressed. The latter interaction is considered within the Hamiltonian ${\cal H}_{int}$, 
whose characteristic oscillator frequency is $\omega_0=k/m$ .
Notice that hereafter the period of the external potential, $b$, is taken as the length scale of the system and the magnitude 
$kb^2$ will be the energy unit.~\cite{Hu98} 

\begin{figure}[ht]
\centerline{\includegraphics[width=70mm,clip]{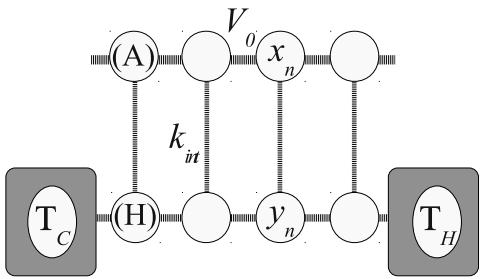}}
\caption{Schematic view of a fragment of the ladder-model under consideration.
The harmonic lattice is connected at left and right edges to two independent heat reservoirs at temperatures $T_C$ and $T_H$ respectively.}
\label{sketch}
\end{figure}

Fig.~\ref{sketch} presents a schematic view of the system under study.

Our aim is to study the formation of a thermal gradient along this double-stranded structure 
by connecting two independent heat reservoirs at both ends of the harmonic lattice.
Heat baths effects will be included in the dynamics of the edge sites of the system, $y_1$ and $y_N$,
by using stochastic Langevin equations.~\cite{Lepri03}  
This will modify the Newton's equations of motion of the affected sites, $n=1$ and $n=N$, as follows:
\begin{equation}
\frac{d^2y_{n}}{dt^2}=-W^{\prime}(y_n,y_{n-1})-W^{\prime}(y_{n},y_{n+1})-\gamma\frac{dy_n}{dt} + f_n(t) \ ,
\label{NewtonLang}
\end{equation}
where $\gamma$ is the coupling between the system and the stochastic bath and the prime indicates derivative
with respect to $y_n$. Random forces, which include temperature effects, are defined as:
\begin{equation}
\langle f_{n}(t) f_{n}(t')\rangle= 2 T \gamma \delta(t-t')\ .
\label{StocForces}
\end{equation}

The relationship between this dimensionless temperature $T$ and the real one $T_r$ 
is given by $T=K_B T^r/k b^2$, where $K_B$ refers to the Boltzmann constant.~\cite{Hu98}
The temperature of the baths $T$ will be different for the left cold ($T_C$) and right hot ($T_H$) sides of the chain.


Exploiting the equations of motion, the local heat flux is defined by the continuity equation as:
\begin{eqnarray}
 J_n&=&\dot{x}_n\Big(\frac{\partial W^{\prime}(x_n,x_{n-1})}{\partial x_n}+k_{int}\frac{\partial W^{\prime}(x_n,y_n)}{\partial x_n}\Big)\nonumber \\
&+&\dot{y}_n\Big(\frac{\partial W^{\prime}(y_n,y_{n-1})}{\partial y_n}+k_{int}\frac{\partial W^{\prime}(x_n,y_n)}{\partial y_n}\Big)\ .
\label{localflux}
\end{eqnarray}

Molecular dynamics simulations are performed by the numerical method proposed by Greenside and Helfand as a correction 
to the Runge-Kutta method for stochastic equations (3o4s2g),~\cite{Helfand79,Greenside81,Casado03} by considering 
a long enough integration time such that the stationary state is established.
The time step is $\delta t=10^{-5}$ and the friction constant of the baths is set to $\gamma=0.5$ in all simulations.
In the final state the time averaged heat flux reaches a constant value along the system such that $J =\langle J_1\rangle=...=\langle J_N\rangle$.
Thus, the thermal conductivity for a finite system can be calculated as $\kappa =J N/(T_C-T_H)$.
Notice that $\kappa$ will be size-independent in the case of normal heat transport.   
Similarly, in the steady state the time averaged temperature calculated as
$T_n=\langle \dot{x}^2_n + \dot{y}^2_n \rangle$ will reach the stationary thermal profile.

\section{Single anharmonic chain}
\label{single}
To better understand the influence of the coupling to nonlinear degrees of freedom, first
we would like to perform a preliminary study of a single Frenkel-Kontorova chain connected to
two independent heat reservoirs at temperatures $T_C=T_{M}-\delta T$ and right $T_H=T_{M}+\delta T$ at the left and right 
edges of the chain. Thus, we integrate the Newton's equations derived from ${\cal H}_{A}$ and heat baths effects are included 
in the dynamics of sites $x_1$ and $x_N$ according to Eq.~\ref{NewtonLang}.

\begin{figure}[ht]
\centerline{\includegraphics[width=80mm,clip]{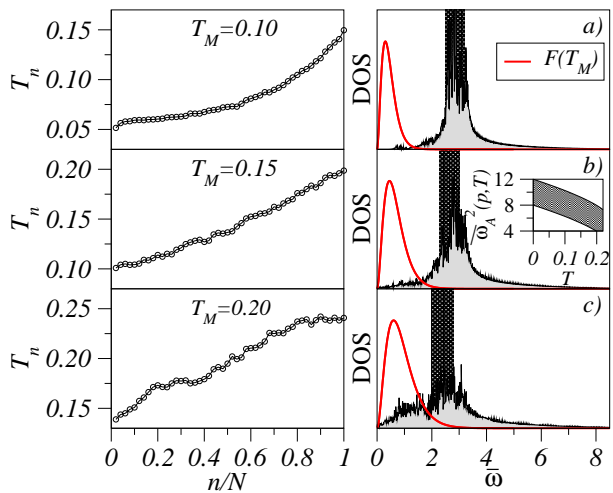}}
\caption{Left panels show the stationary temperature profile calculated by numerical simulations in a FK lattice of $N=50$ sites connected
to two independent heat reservoirs at different temperatures, $T_{H,C}=T_{M}\pm 0.05$. 
Right panels present the comparison between the numerical DOS calculated in equilibrium with the theoretical 
frequency band obtained according to the SCPT (shadowed region). The solid line accounts for the weight function $F(T_{M})$. 
The temperatures considered in the simulations are a) $T_{M}=0.10$, 
b) $T_{M}=0.15$ and c) $T_{M}=0.20$. The inset shows the frequency band of states calculated by SCPT as a function 
of temperature and $V_0=8$ in the shadowed region.}
\label{singleU8}
\end{figure}

We have performed molecular dynamics simulations for different values of $T_{M}$, $\delta T=0.05$ and $V_0=8$.
Once the steady state is reached, the stationary local temperature profile is calculated as $T_n=\langle \dot{x}^2_n\rangle$
which is shown in left panels of Fig.~\ref{singleU8}. 

As it is already known, our results in Fig.~\ref{singleU8} (b) and (c) 
show that if the FK-potential strength is large enough respect to the temperature $T_{M}$, 
a well-defined thermal gradient is formed. Furthermore, in such a case, the Fourier heat law is valid and the total heat flux along the system
does not diverge in the thermodynamic limit contrary to the situation occurring in harmonic systems.
Fig.~\ref{singleU8}(a) demonstrates on the other hand that $T_{M}$ has to be high enough to be able to activate 
the vibrational states of the anharmonic band. This can be understood by invoking a Landauer-like equation to calculate heat 
flux in the ballistic regime.~\cite{Zhong09}
\begin{equation}
J\sim\int \hbar \omega  [n_C(\omega)-n_H(\omega)] \tau(\omega) d\omega.
\label{LandauerFlux}
\end{equation}
Here $\tau(\omega)$ refers to the transmission coefficient, being $\omega$ the frequency of every vibrational state,
and the mode distribution of the heat baths are considered classical within the Maxwell-Boltzmann statistics
$n_{R,L}(\omega)=\exp(-\hbar \omega/K_B T_{R,L}^r)$. In the limit of linear response, $\delta T<<T_{M}$, 
the thermal conductivity can be written as:~\cite{Zimmermann08}
\begin{equation}
\kappa=\frac{J}{2 \delta T}\sim \int x^2 \exp(-x) \tau(x) dx.
\label{LandauerCond}
\end{equation}

The integral in Eq.~\ref{LandauerCond}, being $x=\hbar \omega/K_B T_{M}^r$, is given by the product between the transmission coefficient and 
a weight function $F(T_{M})=x^2 \exp(-x)$,
which accounts for bath thermal effects. This means 
that $F(T_{M})$ defines which vibrational system modes can be excited by the heat reservoirs and contribute to the heat flux.
Right panels of Fig.~\ref{singleU8} show the numerical density of vibrational states (DOS), calculated by Fourier transforming the velocity autocorrelation 
function of equilibrium molecular dynamics simulations, namely, $T_C=T_H=T_{M}$. We also plot the 
weight function $F(T_{M})$ for the $T_{M}$ considered in the simulations.
It is clear that for low $T_M$ there is very few states under the curve 
$F(T_{M})$, and therefore, only those states are involved in heat transfer along the system, see left panel of Fig.~\ref{singleU8}(a). 
The lack of heat carrying modes gives rise to a stationary thermal profile deviated from the linear thermal gradient and thus, 
Fourier's law is not expected to be valid.

To get further insight into the behavior of the nonlinear system the self-consistent phonon theory (SCPT) can be considered.
This approach consists in replacing the anharmonic potential for a harmonic approximation such that the new effective harmonic
strength is temperature-dependent as follows:~\cite{Dauxois93,Shao08} 
\begin{equation}
V(x_n)=\frac{-V_0}{4\pi^2}\cos (2\pi x_n)\rightarrow \frac{U(T)}{2}x_n^2\ , 
\label{SCPTpot}
\end{equation}

By performing a variational study,~\cite{Dauxois93} the effective interaction
$U(T)$ for the Hamiltonian ${\cal H}_{A}$ can be obtained by solving the following transcendental equation:    
\begin{equation}
U(T)=V_0 \exp\Big(\frac{-2 T\pi^2}{ \sqrt{U(T)(U(T)+4)}}\Big)
\label{SCPT}
\end{equation}
Once $U(T)$ is known, the dispersion relation of the system, $\bar \omega_A^2(p,T)= U(T)+ 2 (1-cos(p))$, being $p\in[0,\pi]$, 
and the DOS for the effective harmonic system can be defined analytically. Hereafter $\bar \omega$ refers to frequencies expressed in units of the fundamental frequency $\omega_0=k/m$
of the system.
The frequency band of the SCPT for $V_0=8$ as a function of the temperature is 
presented in the shadowed region of the inset of Fig.~\ref{singleU8}. 
The comparison between the frequency band defined by $\bar \omega_A^2(p,T)$ and the numerical DOS calculated in equilibrium, 
is also shown in the right panel of Fig.~\ref{singleU8} by considering 
the solution of Eq.~\ref{SCPT} at $T=T_{M}$. 

Figure~\ref{singleU8} shows from top to bottom that the anharmonicity shifts the 
DOS to lower frequencies when $T_M$ is increased, as predicted by the SCPT approach. 
However, it is well known that the SCPT approximation is not able to predict the broadening of the frequency 
band when thermal effects become more relevant at higher temperatures, see Fig.~\ref{singleU8}(c).
In view of these arguments, we want to stress two conclusions which will be taken into account hereafter.
On one hand it is clear that $T_M$ in the system should be high enough to activate 
the anharmonic modes in order to create a well formed thermal gradient along the lattice in presence of the thermal baths. 
However, in order to have an accurate description of the thermal properties based on the SCPT, $T_M$ should be low enough 
so that thermal effects are kept within this approach validity.
For the sake of clarity we will focus on temperature regimes which fulfilled these two requirements hereafter.

\section{Ladder-model}
\label{double}
Now that we have reviewed some of the characteristics of an anharmonic one dimensional lattice, 
we focus on the main system of interest in this work, the ladder-model presented in Fig.~\ref{sketch}, 
whose Hamiltonian is described by Eq.~\ref{Hamiltonian}.
Notice that the heat reservoirs at temperatures $T_C$ and $T_H$ are connected to both edges 
of the harmonic chain. Therefore, the dynamics of sites $y_1$ and $y_N$ will be described 
by Eqs.~\ref{NewtonLang} accordingly.

As previously mentioned, a well formed thermal gradient cannot arise along an isolated harmonic chain connected 
to two heat baths at different temperatures. However, its thermal properties are expected to change 
by switching on its interaction with the anharmonic chain in our system. Thus, by
increasing $k_{int}$ the effect of the anharmonicity will be intensified in the whole ladder-system. 
Our results demonstrate that for a large enough interchain interaction, the vibrational spectra of both chains become mixed,
and a linear thermal profile will arise in the system.
This transition is shown in left panels of Fig.~\ref{double} for a system of $N=100$~sites, where the anharmonicity
is $V_0=5$ and two interchain interactions, $k_{int}=0.05$ and $k_{int}=1.00$ are considered.
Figure~\ref{double}(c) also shows that the thermal conductivity along the lattice decays as a function of $k_{int}$.
This is a result of the fact that the contribution from the anharmonic chain to the double-lattice dynamics is stronger for a larger coupling 
and therefore, the thermal resistivity is higher.
\begin{figure}[ht]
\centerline{\includegraphics[width=90mm,clip]{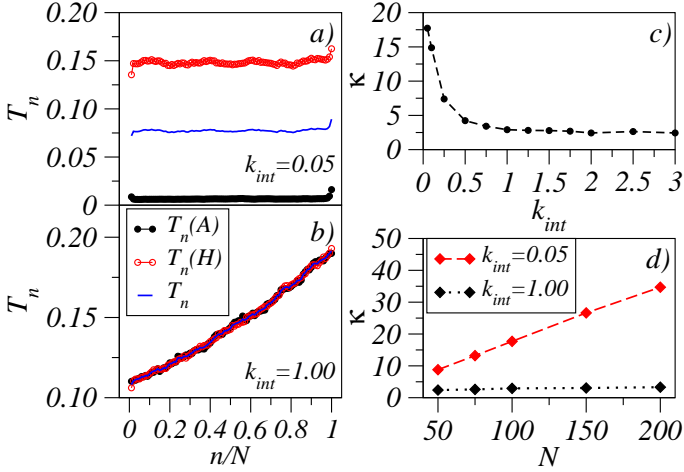}}
\caption{Left panels show the stationary temperature profile calculated by numerical simulations in a ladder-model
(Fig.~\ref{sketch}) of $N=100$ sites connected to two independent heat reservoirs at different temperatures, $T_{C}=0.1$ and $T_{H}=0.2$. 
The temperature profile is plotted for the harmonic $T_n(H)$ and the anharmonic $T_n(A)$ chain, and the coupled system $T_n$.
The considered parameters are $V_0=5$ and a) $k_{int}=0.05$ and b) $k_{int}=1.00$. 
Right panels show the finite-size thermal conductivity as a function of c) the coupling $k_{int}$ and d) the system size $N$ for $k_{int}=0.05$ and $k_{int}=1.00$.}
\label{double} 
\end{figure}
More interestingly, the presence of the thermal gradient
for a large enough coupling is associated to a finite-size thermal conductivity which does not depend on the system size, and 
therefore, Fourier's law is expected to be valid, see Fig.~\ref{double}(d).
   
\begin{figure}[ht]
\centerline{\includegraphics[width=60mm,clip]{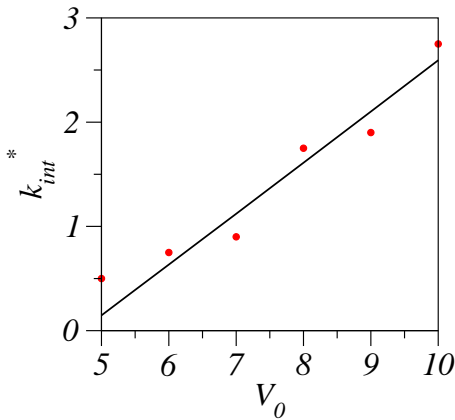}}
\caption{Critical coupling $k_{int}^*$ necessary to create a thermal gradient in the ladder-system
(Fig.~\ref{sketch}) of $N=100$ sites as a function of the anharmonicity $V_0$. Solid line represents the theoretical prediction Eq.~\ref{threshold} in comparison
with the values for $k_{int}^*$ numerically obtained, solid dots. In all cases $T_{C}=0.1$ and $T_{H}=0.2$ have been considered}
\label{critic} 
\end{figure}

In order to give an estimation of the threshold interaction $k_{int}^*$ necessary  
to create a linear thermal gradient in the ladder-model, we will consider a 
fully harmonic ladder-system, whose vibrational spectrum consists of two bands, an acoustic and an optical one.  
In particular the optical band will be at $\bar \omega_H^2(p,k_{int})= 2k_{int} + 2 (1-cos(p))$, namely,
it will be shifted to higher frequencies by increasing the $k_{int}$.
Note that in the real system under consideration (Fig.~\ref{sketch}) for small couplings the anharmonic band will play the 
role of the optical band in the harmonic ladder-model. 
Once the interaction is large enough so that the anharmonic lattice
can be affected by the heat reservoirs and thus, it can contribute to establish the thermal gradient
it will present a band of anharmonic states at frequencies $\bar \omega_A^2(p,T_{M})= U(T_{M})+ 2 (1-cos(p))$. 
This means that, the higher the temperature affecting the anharmonic lattice 
is, the lower the frequencies of the activated anharmonic modes are, 
see Sec.~\ref{single}. 
Therefore, we will define the threshold coupling such that the bands defined by
$\bar \omega_H^2(p,k_{int})$ and $\bar \omega_A^2(p,T_{M})$ will have spectral overlap. 
\begin{figure}[ht]
\centerline{\includegraphics[width=80mm,clip]{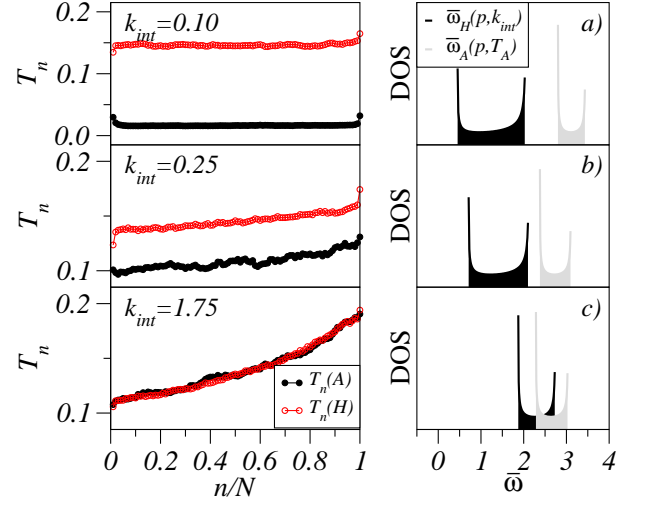}}
\caption{Left panels show the stationary temperature profile for the harmonic $T_n(H)$ and the anharmonic chain $T_n(A)$ 
calculated by numerical simulations in a ladder-model (Fig.~\ref{sketch}) 
of $N=100$ sites connected to two heat reservoirs at $T_{C}=0.1$ and $T_{H}=0.2$. 
Right panels present the theoretical DOS of a single FK chain affected by a mean temperature $T_A$ according to the SCPT,
as well as the DOS of the optical band of a fully harmonic ladder-model for several $k_{int}$.
The considered parameters are $V_0=8$ and a) $k_{int}=0.1$, b) $k_{int}=0.25$ and c) $k_{int}=1.75$.}
 \label{evol}
\end{figure}
According to several simulations, we have concluded that the overlap condition 
between the bands $\bar \omega_H^2(p,k_{int})$ and $\bar \omega_A^2(p,T_{M})$ 
depends on the temperature present in the system.
This is related to the thermal broadening effects mentioned in Sec.~\ref{single}.
When these effects, not included in SCPT, are negligible, we have numerically tested 
that the thermal gradient arises if approximately half of the states are common to the two considered bands.
Thus, the interaction threshold can be estimated by:
\begin{equation}
k_{int}^*=\frac{U(T_{M})-2}{2}\ .
\label{threshold}
\end{equation}
Figure~\ref{evol} shows the formation of the thermal gradient along the ladder-model by increasing $k_{int}$
for a system of $N=100$~sites and $V_0=8$. The right panels present the theoretical DOS corresponding to the band $\bar \omega_A^2(p,T_{A})$ 
of a single anharmonic lattice affected by an averaged temperature $T_{A}=\sum_nT_n(A)/N$ 
obtained in the simulations for each coupling. Additionally, the DOS of the theoretical optical 
band for the fully harmonic ladder-model $\bar \omega_H^2(p,k_{int})$ for the considered couplings is plotted. 
In Fig.~\ref{double}(c), it is shown a well-formed thermal gradient along the system for the coupling $k_{int}^*=1.75$, 
for which the DOS of $\bar \omega_H^2(p,k_{int}^*)$ and $\bar \omega_A^2(p,T_{A})$ overlap in 50\% of their area.
In Fig.~\ref{critic} we show a good agreement between our prediction of $k_{int}^*$ according to Eq.~\ref{threshold} and the threshold 
interaction obtained by numerical simulations for a system size of $N=100$~sites and different values of $V_0$.  

\section{Heat rectification}
\label{rect}

\begin{figure}[ht]
\centerline{\includegraphics[width=60mm,clip]{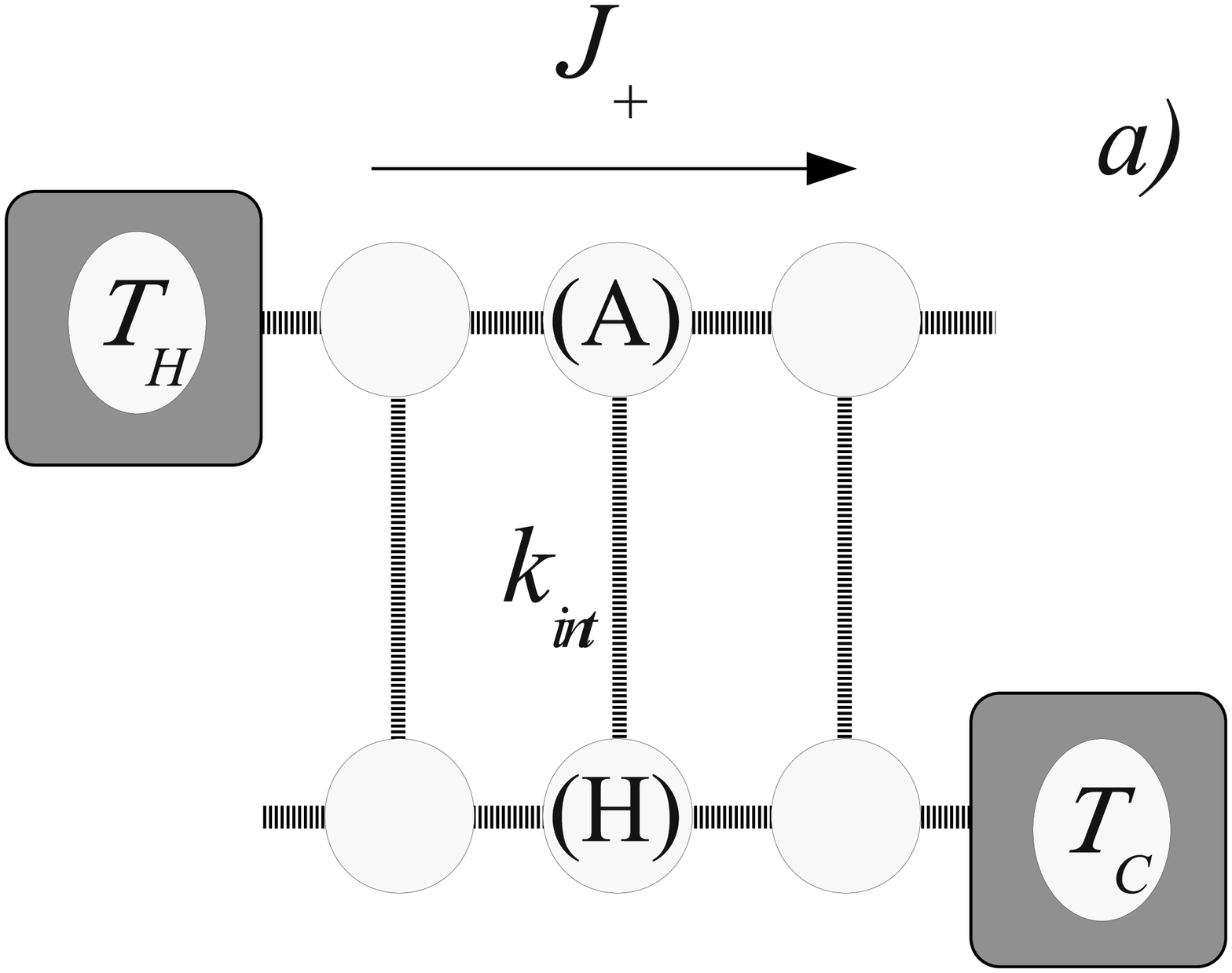}}
\centerline{\includegraphics[width=60mm,clip]{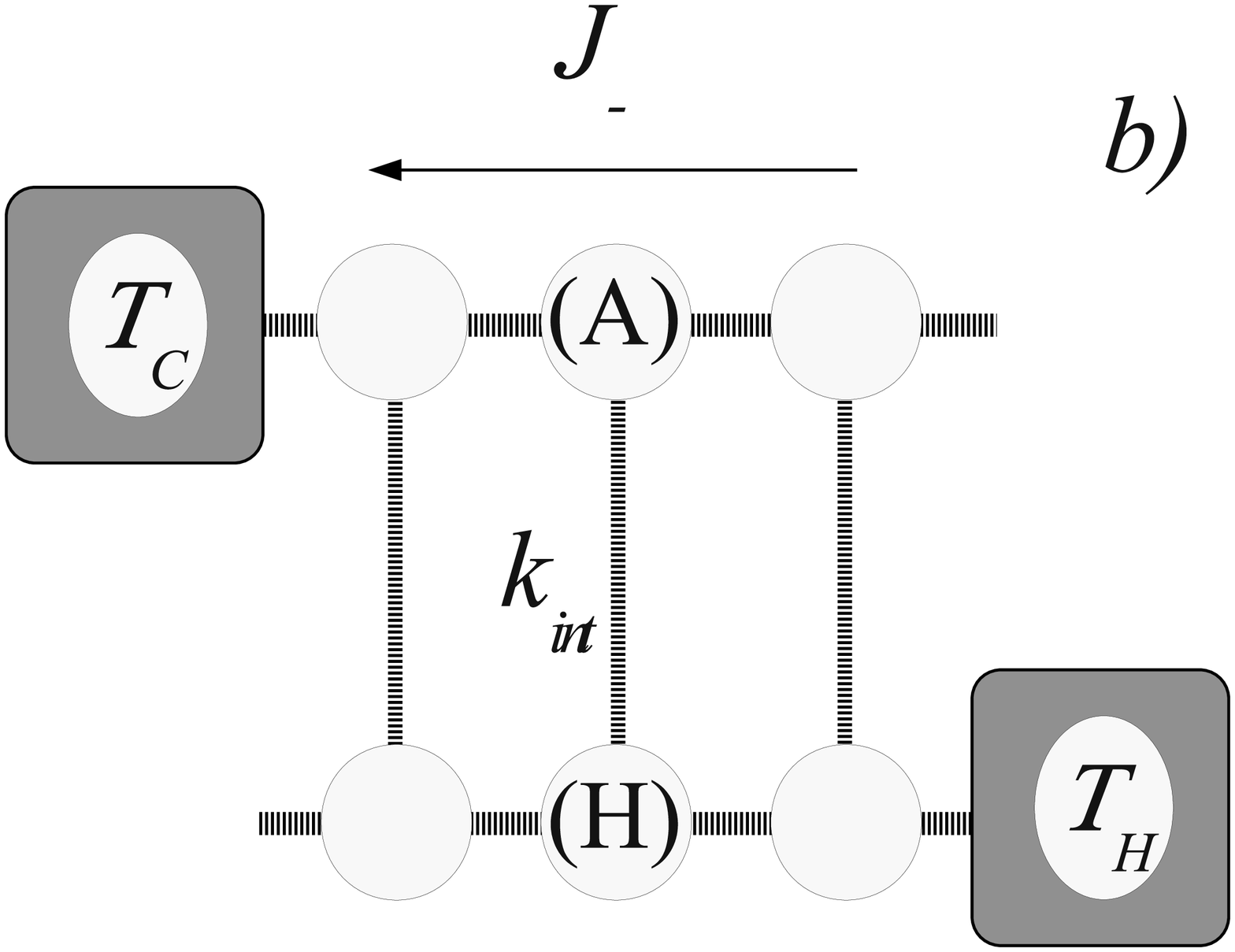}}
\caption{Schematic view of a fragment of the ladder-model under consideration connected to two 
independent heat reservoirs at $T_H$ and $T_C$ in two different configurations. Notice that in a) 
the hot bath is connected to the anharmonic chain and in b) this connection is made to the harmonic one. 
Accordingly heat current flows in opposite directions, $J_+$ and $J_-$, respectively.}
\label{figsrect}
\end{figure}

In previous works it has been well established that symmetry breaking and nonlinearity 
are the necessary ingredients in a system to support heat rectification.~\cite{Terraneo02,Hu06,Li06,Zhang10} In this regard,
we propose a new thermal rectifying mechanism present in the ladder-model under consideration. 
On the one hand, it is clear that nonlinearity comes from the anharmonicity of one of the chains.
On the other hand, the symmetry breaking occurs by considering different connections for the heat reservoirs, 
in a similar way which has been considered in the case of charge transport in DNA.~\cite{RafaelRect}  
Figure~\ref{figsrect} shows the two considered baths configuration where $T_H$ and $T_C$ refer to 
the temperature of the hot and the cold heat reservoirs as previously. 
Notice that in Fig.~\ref{figsrect}(a)
the hot reservoir is connected to the left edge of the anharmonic chain while in Fig.~\ref{figsrect}(b) the position of the 
baths connection is reversed. This means that the heat current flows in opposite directions for both cases
but more interestingly, our results show that these currents reach different values in the stationary state, and therefore the system behaves as a thermal rectifier.

\begin{figure}[ht]
\centerline{\includegraphics[width=90mm,clip]{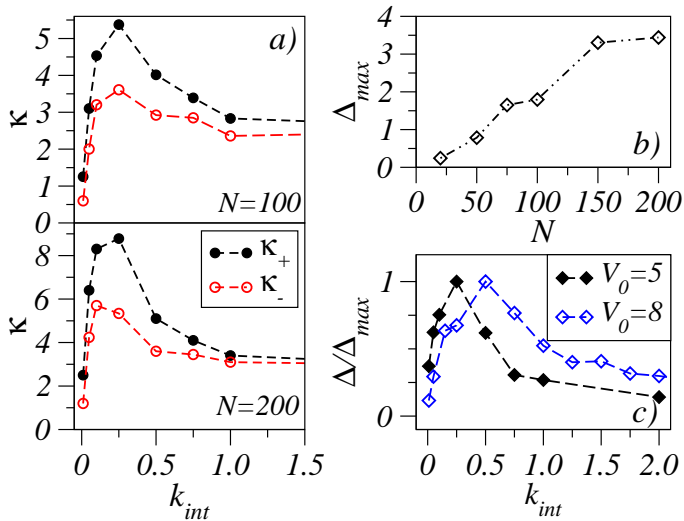}}
\caption{a) Thermal conductivity for the two cases presented in Fig.~\ref{figsrect}, $\kappa_+$ and $\kappa_-$, 
as a function of the interchain coupling with $V_0=5$ and for two different system sizes $N=100$ and $N=200$.
b) System size dependence of the thermal conductivity difference $\Delta_{max}$ for $V_0=5$.
c) Normalized thermal conductivity difference $\Delta/\Delta_{max}$ as a function of the coupling $k_{int}$ for $N=100$ 
and two different anharmonicities $V_0=5$ and $V_0=8$.
}
\label{FigRect} 
\end{figure}

We perform molecular dynamics simulations by connecting the ladder-model described by the Hamiltonian Eq.~\ref{Hamiltonian} to two
heat reservoirs. In Fig.~\ref{figsrect}(a) the dynamics of sites $x_1$ and $y_N$ will be described 
by Eqs.~\ref{NewtonLang} under Langevin baths at temperatures $T_H=0.2$ and $T_C=0.1$ respectively. In Fig.~\ref{figsrect}(b) 
the connections are reversed and sites $y_1$ and $x_N$ are the one affected by Eqs.~\ref{NewtonLang} accordingly.
In order to analysis the main features of the heat rectification, we will numerically calculate the thermal conductivity, 
a) $\kappa_+ =J_+ N/(T_H-T_C)$ and b) $\kappa_- = J_- N/(T_C-T_H)$, for several couplings $k_{int}$. Notice that the interchain interaction has revealed as 
the key parameter to change the thermal properties of the ladder-model in Sec.~\ref{double}. These results are presented in Figure~\ref{FigRect}(a)
for $V_0=5$ and two different system sizes $N=100$ and $N=200$.

Our results establish the existence of heat rectification, $\kappa_+>\kappa_-$, for a certain range of the interchain interaction $k_{int}$.
The difference between both conductivities $\Delta=\kappa_+-\kappa_-$ reaches a maximum $\Delta_{max}$ for a particular coupling   
$k_{int}^{0}\sim 0.25$, where we find a strong rectification effect of 35\%. Qualitatively we can understand this rectifying behavior
by the following handwaving argument. It is clear that if there is no coupling between both chains, the system cannot reach any stationary state,
since the heat current is not able to flow from one bath to the other. However, when the coupling is switched on, 
new channels connecting both chains allow for the heat flow from the hot reservoir to the cold one and the thermal conductivity increases.
Due to the coupling in a) the heat current flows from a bad conductor, the anharmonic chain at $T_H$, 
to a good one, the harmonic chain at $T_C$, while in b) the situation is the opposite one, see Fig.~\ref{figsrect}. 
Therefore, in both cases 
$\kappa_+$ and $\kappa_-$ increases with $k_{int}$ but since the heat current is deviated 
to a more efficient heat conductor in a), $\kappa_+$ is larger than $\kappa_-$. In order to support this argument Fig.~\ref{figsrect}(b) shows that $\Delta_{max}$ increases with 
the system size $N$, this means with the number of coupling channels, as expected. 

Once a good connection between both baths is established, if $k_{int}>k_{int}^{0}$ 
according to Sec.~\ref{double}, the effect of the anharmonicity will be intensified 
in the ladder-system, and therefore the conductivity is expected to decrease, see Fig.~\ref{double}(c). 
When the coupling is so large, that the vibrational spectra of both chains are completely mixed, the rectifying effect 
is expected to disappear and thus $\kappa_+=\kappa_-$. In Fig.~\ref{figsrect}(a) we show that this situation occurs for a $k_{int}^{\infty}\sim1.0$.

Last, we would like to address, how the anharmonicity $V_0$ affects the main features of heat rectification in the system.
For the sake of comparison Fig.~\ref{figsrect}(c) shows the magnitude $\Delta/\Delta_{max}$ as a function of $k_{int}$ for 
two different anharmonicities, $V_0=5$ and $V_0=8$. Our results show that the qualitative rectification behavior does not depend on $V_0$
but its main features are shifted to larger couplings $k_{int}$. In a similar way, in Sec.~\ref{double} it was already demonstrated that the threshold interaction
$k_{int}^*$ for which the vibrational spectra of the subchains are mixed in the system, increases with the anharmonicity $V_0$, see Fig.~\ref{critic}. 
It is to be mentioned that it is not trivial to establish a direct relationship between $k_{int}^{\infty}$ and $k_{int}^{*}$ since the baths connections and 
thus, their thermal effects are different in Fig.~\ref{sketch} and Fig.~\ref{figsrect}. 
However, according to our simulations it turns out that $k_{int}^{\infty}>k_{int}^{*}$, and therefore, we can predict that the main heat rectification effects will appear 
for $k_{int}<k_{int}^{*}$.

\section{Conclusions}
\label{conclusions}

In this paper, we have studied heat conduction along a ladder-model 
consisting of a harmonic and an anharmonic one dimensional chain coupled by harmonic interactions.
We have analyzed how the thermal properties of the coupled ladder-system depends on the strength of the coupling.
In particular, by connecting two independent heat reservoirs to the harmonic counterpart of the system, 
we have demonstrated that for a large enough interchain interaction $k_{int}$, 
a thermal gradient can be formed along the system and the thermal conductivity will remain constant for increasing size systems. 
This means that the ladder-system will present normal heat transport contrary to the case of an isolated harmonic system.
We have estimated the threshold interaction of this transition by considering the self-consistent phonon theory for a single anharmonic chain
in comparison to a fully harmonic ladder-model. Our estimations has been proven to be in good agreement with numerical results 
based on molecular dynamics simulations.

Taking advantage of the nonlinear effects present in our system, it was shown that 
heat current along the ladder-model behaves differently if the heat baths connections are interchanged between the harmonic and the anharmonic chain as 
in Fig.~\ref{figsrect}. This mechanism reveals strong heat rectification effects of more than 30\%. 
The main qualitative features of the rectifying device do not depend  
on the system size nor on the anharmonicity present in the system. However, the maximum rectification rates increases with the system size 
and shifts to larger couplings in the case of stronger anharmonic effects. 
 
\acknowledgments
This work has been supported by DFG-Projekt CU 44/20-1, MCINN-Project MAT2010-17180, and by the South Korea 
Ministry of Education, Science and Technology Program "World Class University" under contract R31-2008-000-10100-0.
E. D. acknowledges financial support by Ministerio de Eduacion y Ciencia and the program Flores Valles-UCM. 
We acknowledge fruitful discussions with H. Sevincli and S. Avdoshenko.


\begin{thebibliography}{99}
\bibitem{Lepri03} S. Lepri, R. Livi, and A. Politi, Phys. Rep. {\bf 377} 1 (2003).
\bibitem{Dhar08} A. Dhar, Adv. Phys. {\bf 57} 457 (2008).

\bibitem{Hu98} B. Hu, B. Li, and H. Zhao, Phys. Rev. E {\bf 57} 2992 (1998).


\bibitem{Gillan85} M. J. Gillan, R. W. Holloway, J. Phys. C {\bf 18} 5705 (1985).
\bibitem{Tsironis99} G. P. Tsironis, A. R. Bishop, A. V. Savin, and A. V. Zolotaryuk, Phys. Rev. E {\bf 60} 6610 (1999).
\bibitem{Li06_fi4} N. Li, P. Tong, and B. Li, Europhys. Lett.  {\bf 75} 49 (2006).
\bibitem{Zhong09} W-R. Zhong, P. Yang, B-Q. Ai, Z-G. Shao, and B. Hu, Phys. Rev. E  {\bf 79} 050103(R) (2009).

\bibitem{Aoki00} K. Aoki, D. Kusnezov, Phys. Lett. A {\bf 265} 250 (2000).
\bibitem{Hu99} B. Hu, B. Li, and H. Zhao, Phys. Rev. E {\bf 61} 3828 (1999).
\bibitem{Li07} N. Li and B. Li, Phys. Rev. E {\bf 76} 011108 (2007).
\bibitem{Piazza09} F. Piazza and S. Lepri, Phys. Rev. B {\bf 79} 094306 (2009).


\bibitem{Terraneo02} M.Terraneo, M. Peyrard, and G. Casati, Phys. Rev. Lett. {\bf 88} 094302 (2008).
\bibitem{Hu06} B. Hu, L. Yang, and Y. Zhang, Phys. Rev. Lett. {\bf 97} 124302 (2006). 
\bibitem{Li06} B. Li, L. Wang, and G. Casati, App. Phys. Lett. {\bf 88} 143501 (2006).
\bibitem{Zhang10} L. Zhang, J-S. Wang, and B. Li, Phys. Rev. B {\bf 81} 100301(R) (2010).

\bibitem{Lee05} L. W. Lee and A. Dhar, Phys. Rev. Lett. {\bf 95} 094302 (2005).
\bibitem{Saito10} K. Saito and A. Dhar, Phys. Rev. Lett. {\bf 104} 040601 (2010).
\bibitem{Liu07} Z. Liu and B. Li, Phys. Rev. E {\bf 76} 051118 (2007).
\bibitem{Liu10} Z. Liu, X. Wu, H. Yang, N. Gupte, and B. Li, New J. Phys. {\bf 12} 023016 (2010).

\bibitem{Iguchi97} K. Iguchi, Int. J. Mod. Phys. B {\bf 11}, 2405 (1997).
\bibitem{Diaz07}  E. D\'{i}az, A. Sedrakyan, D. Sedrakyan, and F. Dom\'{i}nguez-Adame, Phys. Rev. B {\bf 75} 014201 (2007).
\bibitem{Scott92} A. Scott, Phys. Rep. {\bf 217}, 1 (1992). 
\bibitem{Henning02} D. Hennig, Phys. Rev. B {\bf65}, 174302 (2002).
\bibitem{Zhong10} W-R. Zhong, Phys. Rev. E {\bf 81} 061131 (2010).

\bibitem{Dauxois93} T. Dauxois, M. Peyrard, and A. R. Bishop, Phys. Rev. E {\bf 47} 684 (1993).
\bibitem{Shao08} Z-G. Shao, L. Yang, W-R. Zhong, D-H. He, and B. Hu, Phys. Rev. E {\bf 78} 061130 (2008).

\bibitem{Helfand79} E. Helfand, Bell Syst. Tech. J. {\bf 58} 2289 (1979).
\bibitem{Greenside81} H. S. Greenside and E. Helfand, Bell Syst. Tech. J. {\bf 60}, 1927 (1981).
\bibitem{Casado03} J. Casado-Pascual \textit{et al.}, Phys. Rev. E \textbf{67}, 036109 (2003). 

\bibitem{Zimmermann08} J. Zimmermann, P. Pavone, and G. Cuniberti, Phys. Rev. B {\bf 78} 045410 (2008).

\bibitem{RafaelRect} R. Gutierrez, S. Mohapatra, H. Cohen, D. Porath, and G. Cuniberti, Phys. Rev. B {\bf 74} 235105 (2006).

\end{thebibliography}
\end{document}